\documentclass[%
prl, reprint,
superscriptaddress,
 amsmath,amssymb,
 aps,
]{revtex4-1}

\usepackage{graphicx}
\usepackage{epstopdf}
\usepackage{dcolumn}
\usepackage{bm}
\usepackage[usenames, dvipsnames]{color}

\begin{document}

\title{Quantum Simulations with a Trilinear Hamiltonian}
\author{Shiqian Ding}
\altaffiliation{dingshq@gmail.com\\Present address: JILA, National Institute of Standards and Technology and University of Colorado, and Department of Physics, University of Colorado, Boulder, CO 80309, USA}
\author{Gleb Maslennikov}
\author{Roland Habl{\"u}tzel}
\affiliation{Centre for Quantum Technologies, National University of Singapore, 3 Science Dr 2, 117543, Singapore}
\author{Dzmitry Matsukevich}
\affiliation{Centre for Quantum Technologies, National University of Singapore, 3 Science Dr 2, 117543, Singapore}
\affiliation{Department of Physics, National University of Singapore, 2 Science Dr 3, 117551, Singapore}

\date{\today}

\begin{abstract}
Interaction among harmonic oscillators described by a trilinear Hamiltonian $\hbar \xi (a^{\dagger} b c + a b^{\dagger} c^{\dagger}$) is one of the most fundamental models in quantum optics.  By employing the anharmonicity of the Coublomb potential in a linear trapped three-ion crystal, we experimentally implement it among three normal modes of motion in the strong-coupling regime, where the coupling strength is much larger than the decoherence rate of the ions motion. We use it to simulate the interaction of atom and light as described by the Tavis-Cummings model and the process of nondegenerate parametric down conversion in the regime of depleted pump. 
\end{abstract}

\pacs{Valid PACS appear here}

\maketitle

The system of three quantum harmonic oscillators where a quantum in a mode is down-converted into two quanta in other two modes is a cornerstone model in many branches of physics \cite{firstparametric1961quantum}, including quantum optics and quantum information science~\cite{review2005RMP,2016RMP}. It describes physical processes ranging from Raman or Brillouin scattering \cite{Raman1969Walls}, nondegenerate parametric down-conversion of light in nonlinear medium~\cite{Mandell_book}, operation of absorption refrigerator~\cite{ourabsorptionpaper,quantum_fridge_2012,quantum_enhanced_fridge_2014,2015singleshotabsorption}, to zero-dimensional model of Hawking radiation~\cite{2010_hawking_trilinear}, and is equivalent to the widely used Tavis-Cummings model~\cite{dickemodel1954,tavismodel1968, walls_1970, carusotto_1989}, which describes the interaction of radiation field with $N$ two-level atoms. 

The weak coupling between the modes of light can be easily achieved with nonlinear crystals 
and is commonly used, for example, to generate entangled photon pairs~\cite{Mandell_book}. 
It is widely exploited in the experiments such as demonstration of the Einstein-Podolsky-Rosen paradox \cite{1992EPR_Kimble}, generation of the two-mode squeezing \cite{2014PRL_twomodesqueezing}, and realization of the frequency-entangled photon pairs spectroscopy \cite{2003APL_biphotonspec,2004PRA_biphotonspec,2016RMP}. The nondegenerate parametric interaction has also been recently realized in the electro/opto-mechanical systems \cite{2011Nature_microampl,2013Science_Lehnert,2014PRL_Japan,2015PRL_twomodesqueezing,2016Nature_nonclassical,2017PRL_nanomechanical}, and the superconducting circuits \cite{2010Nature_superconducting,2012PRL_Devoret,2011PRL_Wallraff,2012PRL_Huard2,2015PRL_Huard}.
An analogous interaction in hybrid light-matter quasiparticles is proposed to convert a Raman laser into an optical parametric oscillator and create an all-optical switch \cite{2016PRL_RamantoOPO}. Currently, a Raman-like coupling among three harmonic oscillators is demonstrated \cite{Ramanlike_2018} in the graphene membranes via the tension-mediated nonlinear interaction \cite{graphene_2016,graphene_2016_2}.
However, strong interaction at the level of single quanta
remains a challenge, and is required for application in quantum computation
and quantum simulations \cite{review2005RMP}.

In this Letter, we experimentally demonstrate the nondegenerate parametric interaction between mechanical modes of motion in the system of trapped ions 
in the fully quantum regime. 
We consider three identical ions of mass $m$ and electric charge $e$ that are
aligned along the axial $z$ direction of a rf-Paul trap with the single-ion trapping frequencies $\omega_x, \omega_y, \omega_z$.
The size of the trap ($\sim$ mm) is much larger than the length scale of the ion crystal ($\sim \mu$m), such that the trap is well approximated by a harmonic potential \cite{wineland_review}. 
Considering also the Coulomb interaction between the ions, we can write the total potential energy as
\begin{equation}
\label{eq:interaction}
V = \frac{m}{2}\sum_{n=1}^3( \omega_x^2 x_i^2 + \omega_y^2 y_i^2 + \omega_z^2 z_i^2 ) 
+ \sum_{\substack{n,k=1 \\ n \neq k}}^3 \frac{e^2}{8 \pi \epsilon_0 | \vec{r}_n - \vec{r}_k |},
\end{equation}
where 
$\vec{r}_i = (x_i, y_i, z_i)$ denotes the position of the ion $i$, and $\epsilon_0$ is the dielectric permittivity of vacuum. 

\begin{figure*}[tbp!]
\centering
\includegraphics[width=1.75\columnwidth]{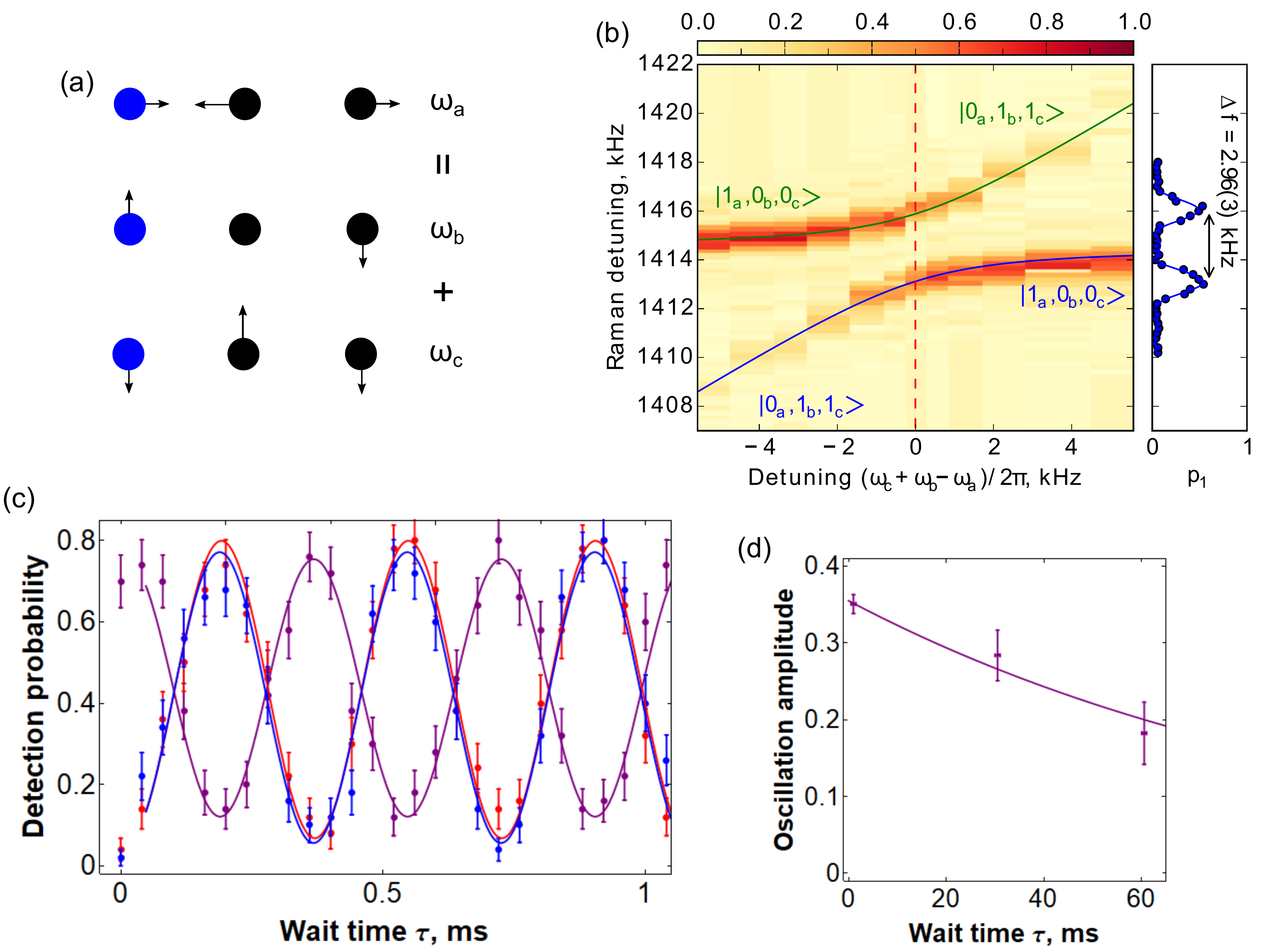}
\caption{\label{fig:coupling}
Trilinear mode coupling. (a) One axial and two radial modes of motion involved in the experiment. Two ions (black) are optically pumped into the metastable $^2F_{7/2}$ state and remain there during the experiment. 
(b) Probability to drive an axial-mode blue sideband transition as a function of Raman detuning (vertical axis) and three-mode detuning $\delta$ (horizontal axis), after all the three motional modes are prepared in the ground state. 
Solid lines are the eigenvalues of the Hamiltonian Eq.(\ref{eq:Hamiltonian}).
Inset: The on-resonance vacuum Rabi splitting measured in the spectrum of the axial-mode blue sideband.
(c) The coherent energy exchange between three modes of motion. The purple, red and blue dots represent the probabilities to drive the corresponding red sideband transitions of the axial zig-zag ($\omega_a=1414$ kHz), the radial tilt ($\omega_b=878$ kHz) and the radial zig-zag ($\omega_c=536$ kHz) modes, respectively, as functions of three-mode interaction time $\tau$. The purple, red and blue lines are the corresponding sinusoidal or cosinusoidal fits.
(d) The oscillations decay with a time constant of $0.11(2)$ s, which corresponds to more than 300 cycles of energy exchange.}
\end{figure*}

Motion of three ions in the trap in harmonic approximation is decomposed into a set of decoupled normal modes: the motion along each trap axis is described by the center-of-mass mode with the eigenvector $e_{cm} = (1, 1, 1) / \sqrt{3}$, the "tilt" mode $e_{t} = (-1, 0, 1) / \sqrt{2}$, 
and the "zigzag" mode $e_{z} = (-1, 2, -1) / \sqrt{6}$. However, anharmonicity of the Coulomb interaction Eq.~(\ref{eq:interaction}) can be significant even 
at the spatial extent of the single-ion wavepacket in the motional ground state ($\sim$ 10 nm), which induces nonlinear coupling between the normal modes for the oscillator energies on the order of single quanta.  

Symmetry and energy conservation restrict the set of modes that can be coupled.
The center-of-mass modes can not be coupled to any other modes, since they are purely determined by the trap potential and do not depend on the Coloumb interaction between the ions. 
We fulfill the energy conservation condition by particularly tuning the mode frequencies to 
\begin{equation}
\label{eq:resonancecondition}
\omega_a = \omega_b + \omega_c,
\end{equation}
where $\omega_b\neq\omega_c$.
When $\omega_z = 0.556\,\omega_x$, this condition is satisfied for the axial zigzag mode with the frequency 
$\omega_a = \sqrt{29/5}\,\omega_z$, and the tilt and the zigzag modes along
the $x$-radial direction with the frequencies $\omega_{b} = \sqrt{\omega_x^2 - \omega_z^2}$ and $\omega_c = \sqrt{\omega_x^2 - 12\,\omega_z^2 / 5}$, respectively [see Fig.~\ref{fig:coupling}(a)]. 

After rewriting Eq.~\eqref{eq:interaction} in the normal-mode coordinates, we obtain the Hamiltonian using the rotating wave approximation near the resonance condition Eq.~\eqref{eq:resonancecondition} as \cite{james_complete_2003}
\begin{equation}
H = \hbar\,\omega_a a^{\dagger} a+\hbar \omega_b\, b^{\dagger} b
+\hbar \omega_c\, c^{\dagger} c+\hbar \xi (a^{\dagger} b c + a b^{\dagger} c^{\dagger}).
\label{eq:Hamiltonian}
\end{equation}
Here, $a$ ($a^\dagger$), $b$ ($b^\dagger$) and $c$ ($c^\dagger$) are the annihilation (creation) operators for the corresponding normal modes of motion, $\xi = 9 \omega_z^2 
\sqrt{\hbar/m \omega_a \omega_b \omega_c}/5 z_0 $ is the coupling rate, and 
$z_0 = (5 e^2 / 16 \pi \epsilon_0 m \omega_z^2)^{1/3}$ is the distance between neighboring ions.

In our experimental setup, three ytterbium ions $^{171}$Yb$^+$ are trapped in a standard
four-rod Paul trap \cite{ourpaper_microwave,ourpaper_parametricoscillator,ourpaper_crosskerr} with the single-ion trapping frequencies $(\omega_x,\omega_y,\omega_z)=2\pi\times(1056,976,587)$ kHz.
Two of the ions are optically pumped \cite{ourpaper_parametricoscillator} into the metastable $^2F_{7/2}$ state with lifetime of about 5 years \cite{Flifetime2000PRA}. 
Due to collisions with the background gas (the vacuum pressure is about $\sim 2 \cdot 10^{-11}$ mbar), the experimentally measured lifetime of the $^2F_{7/2}$ state in our setup is around 15 minutes, long enough to eliminate the interaction between these dark ions and the laser beams throughout a single experiment ($\sim 100$ ms).
The remaining bright ion is positioned at the edge of the crystal for the state preparation and detection of the collective motional modes \cite{2007PRA,ourpaper_crosskerr}, and its position is monitored with an electron-multiplying CCD camera. Once we find the bright ion in the middle of the crystal, we interrupt the rf signal sent to the trap for a few microseconds to let the ion crystal melt and recrystallize. We repeat this process until the bright ion moves to the edge of the crystal \cite{ourpaper_crosskerr}.

Each of the motional modes is initialized to the ground state ($>95\%$ occupancy)
by Doppler cooling followed by sideband cooling at the detuning ($\delta = \omega_a - \omega_b - \omega_c = -2\pi \times 44~\text{kHz}$) that is much larger than the coupling strength $\xi$.
The sideband cooling is achieved by driving the frequency-comb-assisted Raman transitions~\cite{Hayes_2010,ourpaper_microwave,ourpaper_parametricoscillator,ourpaper_crosskerr} via the ion internal hyperfine states $|^2S_{1/2}, F=1, m_F=0\rangle \equiv |{\uparrow} \rangle$  and 
$|^2S_{1/2}, F=0, m_F=0\rangle \equiv |{\downarrow} \rangle$.

To detect the state of the motional mode, we couple it to the ion internal state by driving red and blue motional sidebands \cite{wineland_review}, and the internal state of the ion is then detected using the standard fluorescence techniques \cite{2007PRA}.
All the modes are independently addressed by two pairs of Raman beams \cite{ourpaper_crosskerr}. The Raman beams, each with repetition rate of 76.2 MHz and average power of 80 mW at central wavelength of 374 nm, are generated by frequency doubling a mode-locked picosecond Ti:sapphire laser \cite{ourpaper_microwave}. 

The interaction term of the Hamiltonian in Eq. (\ref{eq:Hamiltonian}) mixes the bare energy eigenstates and gives rise to the vacuum Rabi splitting.
To verify this, after all the motional modes are cooled to the ground state, we bring the mode detuning $\delta$ to zero in around $20~\mu s$ by applying a bias voltage to the the diagonally opposite electrodes of the ion trap \cite{ourpaper_parametricoscillator}. 
The sweeping speed is chosen to be much faster than the coupling rate $\xi$, such that the phonons in the axial and radial motional modes are well defined during the sweeping, but slower than the mode frequencies $\omega_i$ so that that no additional motional excitations are observed.
We then probe the blue sideband of the axial mode,
and find two peaks with equal linewidths, as shown in the inset of Fig.~\ref{fig:coupling}(b).
The splitting $\Delta\omega_a= 2\pi\times2.96(3)~\text{kHz}$ reveals the coupling rate and it is close to the theoretically expected value $2 \langle 011 | H | 100 \rangle / 2 \pi \hbar = \xi/ \pi=2.767$ kHz. The discrepancy can be attributed to the uncertainty 
of the detuning  $\delta$. 
Further detuning scan
near the resonance condition Eq.~(\ref{eq:resonancecondition}) resolves an avoided crossing, as shown in Fig.~\ref{fig:coupling}(b). 

Another signature of the cross-mode coupling is the coherent energy exchange between different modes under the resonance condition. 
We start with one phonon in the axial mode and zero phonon in both radial modes, and then adjust the detuning $\delta=0$ (resonance).
After time $\tau$, we tune the detuning back to its initial value $\delta=-2 \pi \times 44~\text{kHz}$ and measure the phonon populations in all the three modes.
We observe a coherent energy exchange between the axial and the two 
radial modes with the frequency $2.801(2)~\text{kHz}$ [see Fig.~\ref{fig:coupling}(c)], which is consistent with the theory. The amplitude of oscillations reduces to $1/e$ on a time scale of 0.11(2) s, as shown in Fig.~\ref{fig:coupling}(d), which corresponds to about 330 oscillation cycles. 

\begin{figure}[tb]
\centering
\includegraphics[width=\columnwidth]
{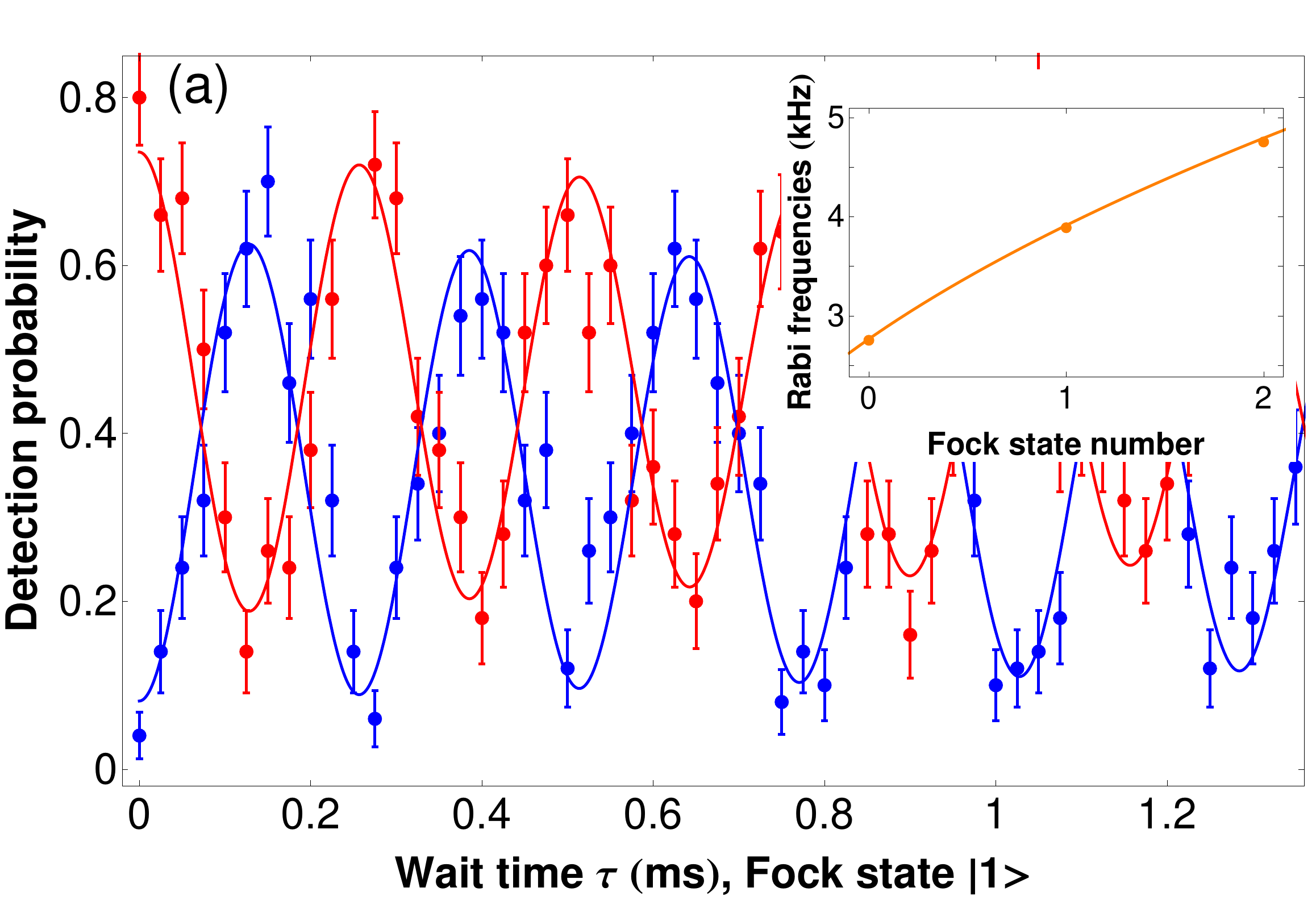}
\includegraphics[width=\columnwidth]
{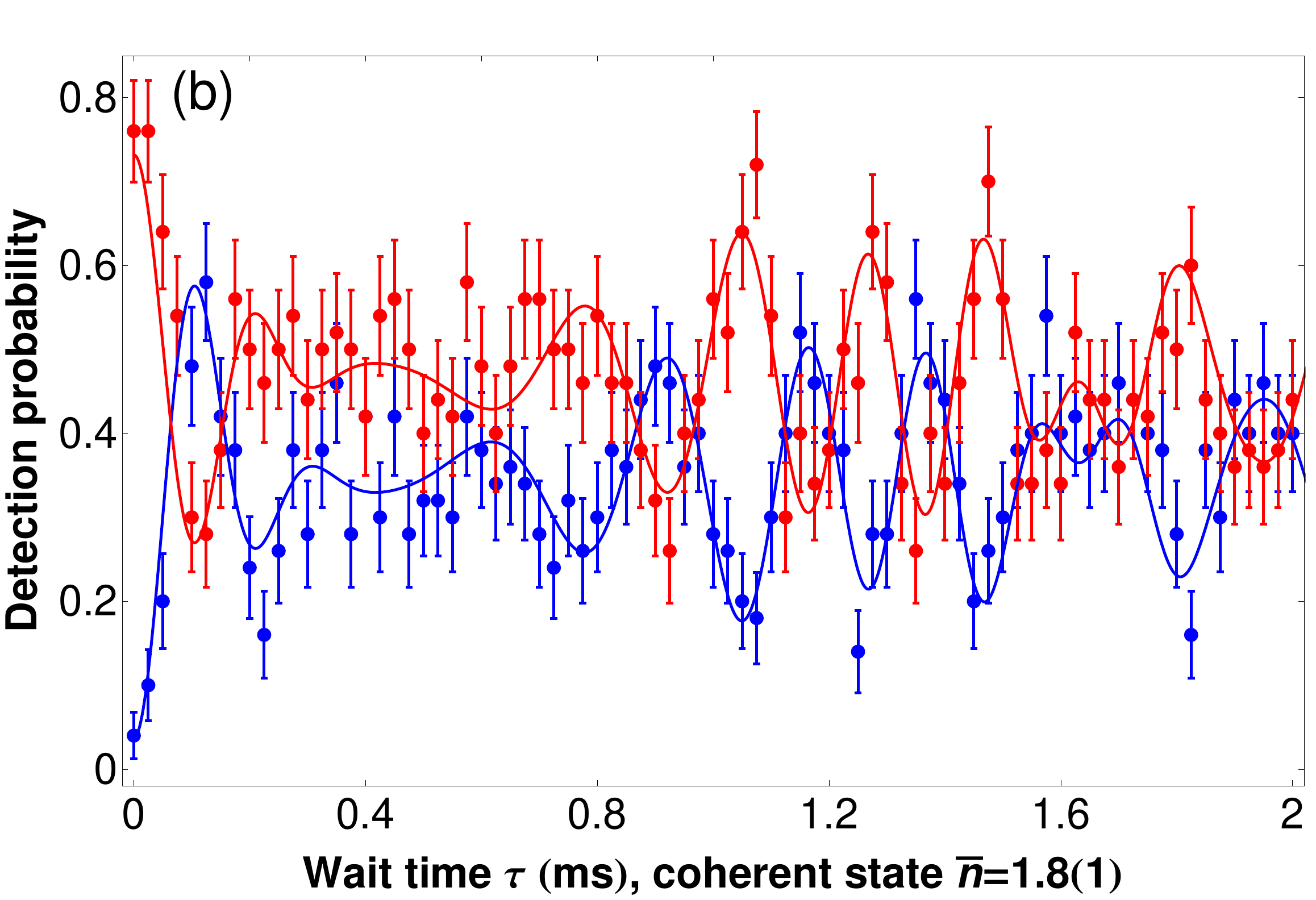}
\caption{\label{fig:Jaynescummingsmodel}
Simulations of the Tavis-Cummings model. Before the system is brought into resonance for interaction, the axial zig-zag and the radial tilt modes are initially both prepared in vacuum state and the radial zig-zag mode in (a) the Fock state $|1\rangle$ and (b) the coherent state with the average phonon number $\bar{n}=1.8(1)$.
The red (blue) dots represent the probability of success to drive the red sideband transitions of the axial zig-zag (radial tilt) mode as a function of interaction time $\tau$, while the red (blue) lines show the corresponding fits.
The inset of Fig. (a) shows the extracted Rabi oscillation frequency from the fits as a function of the phonon numbers in the radial zig-zag mode Fock states. The orange dots are the experimental data and the orange line is the theoretical prediction $\Omega_n/2\pi=2\sqrt{n+1}\xi/2\pi$. 
The extracted $\bar{n}=1.8(1)$ in Fig. (b) from the fits, assuming a Poisson phonon number distribution, is consistent with an independent phonon number calibration \cite{JCions1996PRL,ourpaper_crosskerr}.
The oscillations in this figure correspond to the Rabi oscillations of the initially excited atom interacting with a light field (see the main text).
}
\end{figure}

The Hamiltonian in Eq.~(\ref{eq:Hamiltonian}) in the strong coupling regime has a rich structure that allows simulation of diverse classes of physical phenomena. By introducing the Schwinger's oscillator scheme~\cite{Schwinger1965, sakuraibook}, $J_z=(a^{\dagger}a-bb^{\dagger})/2$, $J_-=ab^{\dagger}$ and $J_+=a^{\dagger}b$, this Hamiltonian can be rewritten as 
\cite{walls_1970,carusotto_1989,chumakov1996dicke}
\begin{equation}
\label{eq:tavis_cummings}
H=\hbar \omega_c (c^{\dagger}c+J_z)+\hbar \xi(c^{\dagger}J_-+cJ_+).
\end{equation}
This Hamiltonian is formally equivalent to the Tavis-Cummings model \cite{tavismodel1968}, which describes the coherent 
interaction of a quantized single-mode light field $c$ with an ensemble of identical spin 1/2 atoms. Here we use $n_a=a^{\dagger}a$ and $n_b=b^{\dagger}b$, i.e., the phonon numbers in the axial zig-zag and the radial tilt modes, to simulate the numbers of spin-up and spin-down atoms respectively, and $J$ is operator of the collective spin of the atoms.

As a proof of principle, we simulate the Jaynes-Cummings model, which is a specific case of the Tavis-Cummings model. 
We prepare one-phonon state in mode $a$ and vacuum state in mode $b$, which corresponds to 
a system consisting of one
spin 1/2 atom in the spin-up state, while the mode $c$ that simulates the quantized light field is prepared in either Fock or coherent states. 
After letting the system  evolve for time $\tau$, we measure the final phonon populations in the 
modes $a$ and $b$. The result is shown in Fig.\ref{fig:Jaynescummingsmodel}. 
If the mode $c$ is initially prepared in the Fock states, 
we observe the periodic phonon population oscillations in both modes $a$ and $b$ (see Fig.\ref{fig:Jaynescummingsmodel}(a)), which corresponds to the Rabi oscillations of the atom interacting with the Fock-state light mode. 
The frequencies of the oscillations increase with the phonon number in $c$ mode as $\Omega_n/2\pi=2\sqrt{n+1}\xi/2\pi$, as shown in the inset of Fig.\ref{fig:Jaynescummingsmodel}(a). 
If the field mode is initially prepared in a coherent state with the population distribution $p_n = e^{-\bar{n}} \bar{n}^{n}/n!$, it leads to the collapses and revivals of the atomic Rabi oscillations as shown in Fig.\ref{fig:Jaynescummingsmodel}(b). In both cases [Fig.\ref{fig:Jaynescummingsmodel}(a) and (b)], the phonon numbers in the axial zig-zag and the radial tilt modes are anticorrelated because the operator for the number of atoms $N=n_a+n_b$ commutes with the total Hamiltonian Eq. (\ref{eq:Hamiltonian}). 

It is worth mentioning that the coherence time of the simulation of the Jaynes-Cummings model here is much longer than that previously demonstrated in cavity QED~\cite{JCcavity1996PRL} or in trapped-ions system using the spin-motion coupling~\cite{JCions1996PRL}. 
Together with the rich toolboxes of preparing and controlling both the internal and the motional states of the trapped ions, especially the capability to deterministically prepare motional states with high number of phonons, the established correspondence offers platform to simulate other aspects of the Tavis-Cummings model, including, for example, the preparation of the superradiant state \cite{dickemodel1954,walls_1970}.

\begin{figure}[tbp!] 
\centering
\includegraphics[width= \columnwidth]{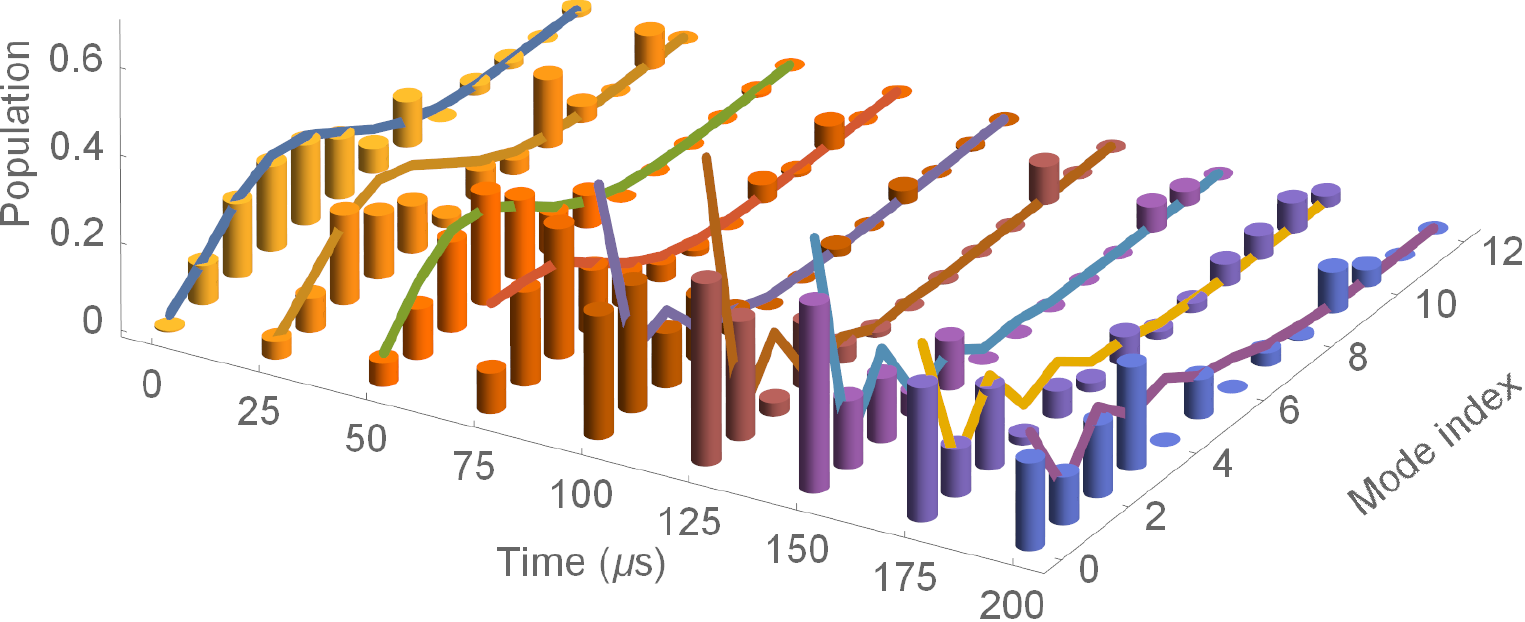}
\includegraphics[width= \columnwidth]{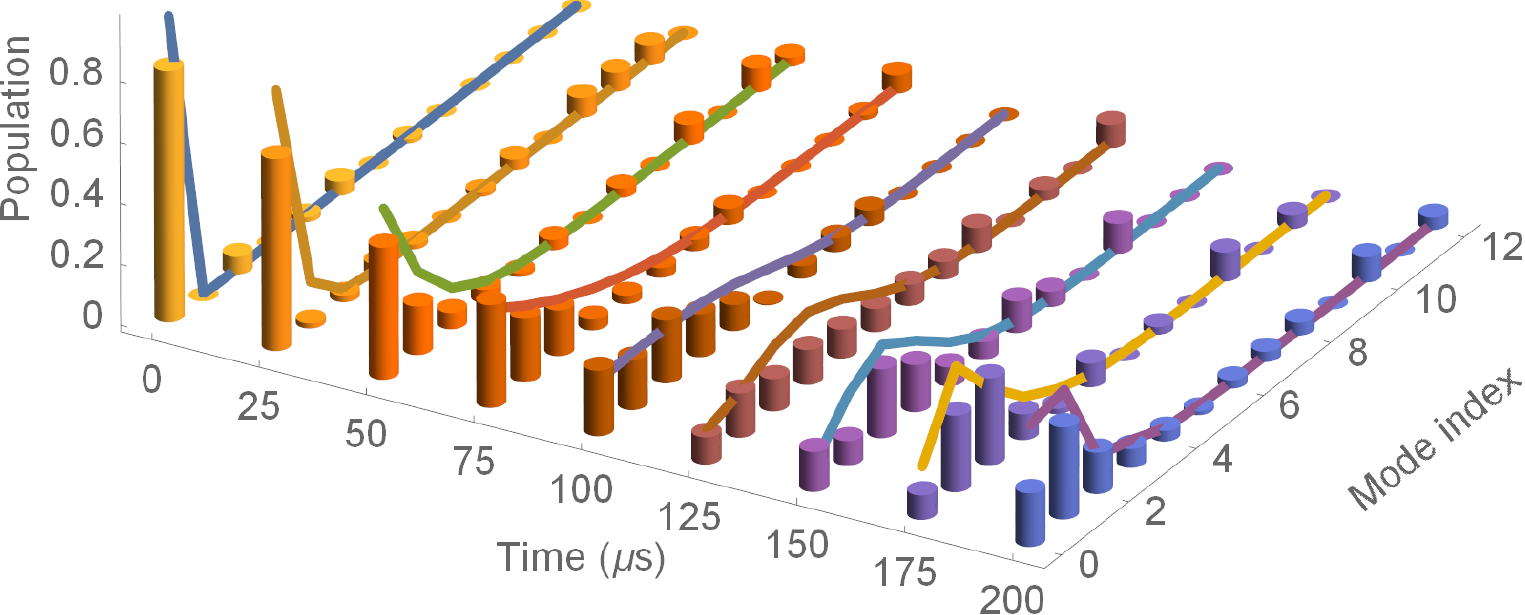}
\includegraphics[width= \columnwidth]{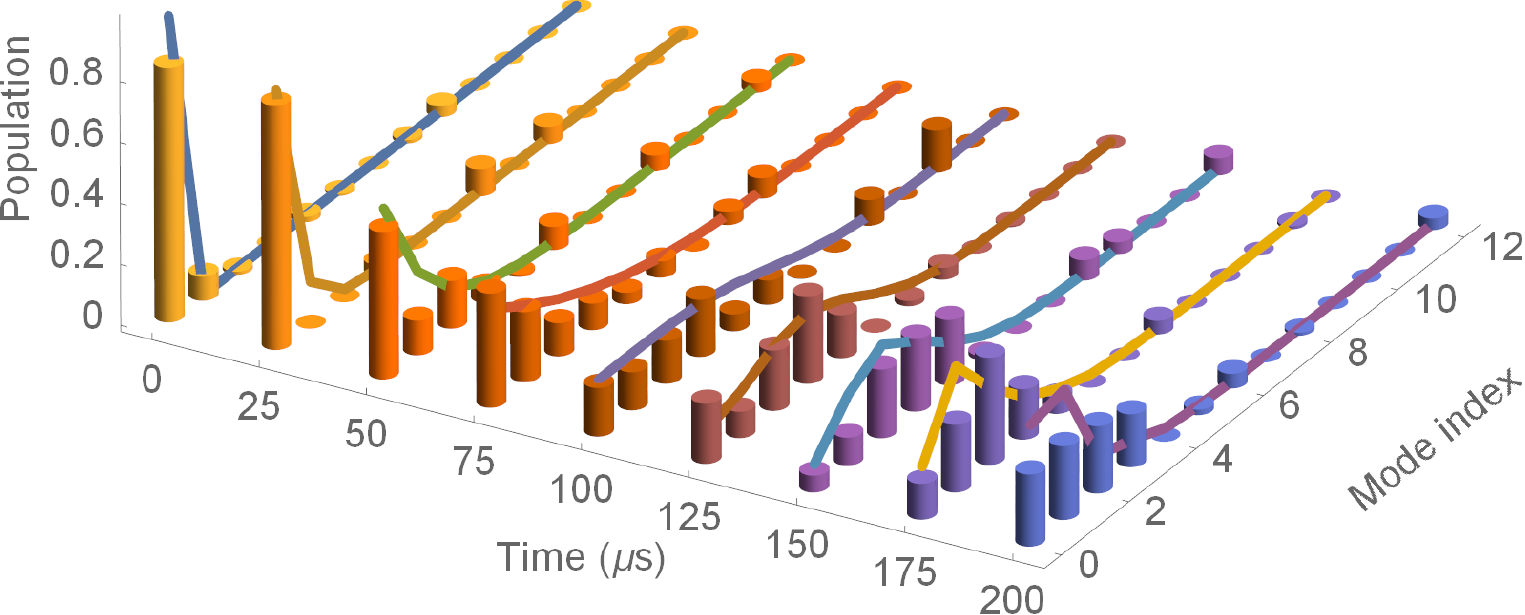}
\caption{\label{fig:oscillations}
Simulation of the nondegenerate parametric down conversion in the depleted-pump regime. The upper, middle, and lower figure panels show the temporal evolutions of the reconstructed phonon number distributions (see the main text) of the axial zig-zag, the radial tilt, and the radial zig-zag modes, respectively. The solid lines are the calculated phonon number distributions 
using Hamiltonian Eq. (\ref{eq:Hamiltonian}) with the axial zig-zag mode initially in coherent state with $\bar{n}=3.7$ and the radial tilt and the radial zig-zag modes both in vacuum state.
} 
\end{figure}

Another physical process that can be simulated with the Hamiltonian [Eq.(\ref{eq:Hamiltonian})] is the nondegenerate parametric down conversion.  
Due to the long coherence time of the motional states and the strong nonlinearity in the trapped-ion system, we are able to simulate this process
in the regime of depleted pump.
We prepare mode $a$ in the coherent state with 
$\bar{n} = 3.7(2)$ and two other modes $b$ and $c$ in the vacuum state.
After interaction time $\tau$,
we drive the blue motional sideband of each of the three motional modes and reconstruct the corresponding phonon number distribution by performing the Fourier transform of the ion internal state temporal evolution \cite{JCions1996PRL,wineland_review}. 
The reconstructed phonon number distributions in each mode for different interaction times are shown in Fig.\ref{fig:oscillations}. If the interaction time is small, the phonon number 
distributions in the modes $b$ and $c$ resemble that of a thermal state, similar to the weak nonlinearity with a strong pump \cite{Mandell_book}. As the interaction time increases, the states of the modes $b$ and $c$ 
significantly deviates from the thermal state. 

Such behavior was also predicted in Ref.\cite{2010_hawking_trilinear} in a simple zero-dimensional model of the Hawking radiation \cite{1974Nature_Hawking}.  
The black hole, represented by mode $a$ here, releases pairs of particles on opposite sides of the event horizon.
One particle escapes from the black hole while the other falls back, emulated by mode $b$ and $c$, respectively.
Initially, the emitted mode obeys thermal statistics. Our observation of deviation from thermal statistics latter
lends credence to the view that the Hawking radiation might be entangled with the quantum 
states of the black hole, and contain information.

To conclude, we demonstrate the trilinear interaction of harmonic oscillators
with a trapped-ion system at the single quanta level,
and use it to simulate the Tavis-Cummings model, and the nondegenerate optical parametric down-conversion. The latter also sheds some light on the information-loss problem of the black hole. 
It offers a broad platform for the field of quantum thermodynamics, including, for example, the simulation of the absorption refrigerator \cite{ourabsorptionpaper,quantum_fridge_2012,quantum_enhanced_fridge_2014,2015singleshotabsorption}, 
and the study of the role of entanglement and the emergence of quantum statistical behavior in an isolated few-body system \cite{Greiner2016,PRL2016thermolization_ion}. 

\begin{acknowledgments} 
This research is supported by the National Research Foundation, Prime Ministers Office, Singapore, and the Ministry of Education, Singapore, under the Research Centers of Excellence program and Education Academic Research Fund Tier 2 (Grant No. MOE2016-T2-1-141).
\end{acknowledgments}

\bibliography{./bib_phonons}
\end{document}